\documentclass[12pt]{article}

\usepackage{amssymb}
\usepackage{amsmath}
\usepackage{amscd}
\usepackage{latexsym}
\usepackage{graphicx}

\usepackage{cite}

\newcommand{\be}{\begin{equation}}
\newcommand{\ee}{\end{equation}}
\newcommand{\Dlt}{\Delta}

\newcommand{\ra}{\rightarrow}

\newcommand{\bt}{\beta}
\newcommand{\vp}{\varphi}

\newcommand{\al}{\alpha}
\newcommand{\sgm}{\sigma}

\newcommand{\Gm}{\Gamma}

\begin{document}

\begin{center}

{\Large {\bf Methods of Retrieving Large-Variable Exponents} \\ [5mm]

V. I. Yukalov $^{1,2}$ and S. Gluzman $^3$} \\ [3mm]

{\it $^{1}$Bogolubov Laboratory of Theoretical Physics, \\
Joint Institute for Nuclear Research,  141980 Dubna, Russia\\ 
{\bf E-mail}: yukalov@theor.jinr.ru \\ [2mm]

$^2$Instituto de Fisica de S\~ao Carlos, Universidade de S\~ao Paulo, \\ 
CP 369, S\~ao Carlos, S\~ao Paulo 13560-970, Brazil \\ [3mm]

$^{3}$Materialica + Research Group, Bathurst St. 3000, Apt. 606, \\
Toronto, ON M6B 3B4, Canada \\
{\bf E-mail}: simongluzmannew@gmail.com or gluz@sympatico.ca }

\end{center}

\vskip 2cm

\begin{abstract}
Methods of determining, from small-variable asymptotic expansions, the
characteristic exponents for variables tending to infinity are analyzed. The
following methods are considered: diff-log Pad\'e summation, self-similar factor
approximation, self-similar diff-log summation, self-similar Borel summation,
and self-similar Borel--Leroy summation. Several typical problems are treated.
The comparison of the results shows that all these methods provide close estimates
for the large-variable exponents. The reliable estimates are obtained when different
methods of summation are compatible with each other.
\end{abstract}

\vskip 2cm
{\parindent=0pt
{\bf keywords}: small-variable asymptotic expansions; large-variable exponents;
diff-log Pad\'{e} summation; self-similar factor approximation; self-similar
diff-log summation; self-similar Borel summation }

\newpage

\section{Introduction}

In many physics problems, one needs to find out the behavior of a function at large
variables tending to infinity. However this function is defined by such complicated
equations that it can be found only as an asymptotic expansion for small variables.
Then the problem arises: How is it possible to extrapolate the small-variable expansion
to the large values of the variable, and even to the variable tending to infinity?

The class of functions that exhibits power-law behavior at large variables is quite
wide. The problem of finding out the large-variable behavior of power-law functions
happens in many applications, where the most important point is to characterize the
type of the power law, as the related characteristic exponent sheds light on the
physical processes responsible for the particular asymptotic behavior. The typical
example is the determination of critical exponents at phase transitions. This problem
is known to be straightforwardly reducible, by the change of variables, to the
definition of the characteristic exponent at infinity \cite{Sornette_2003}.

Another well-known problem is the determination of the tail characteristic exponents
of distributions exhibiting power laws, such as the Pareto law \cite{Pareto_1927}
and Zipf law \cite{Saichev_2010}. The character of the large-variable behavior of a
distribution describes the type of the variable mean, and its variance, which portray
the properties of the considered system \cite{Sornette_2003,Saichev_2010}.

Important information on the properties of many-body systems, e.g., on spatial structure,
collective excitations, and impenetrable obstacles can be derived from the study of the
large-variable tails in scattering theory \cite{Migdal_2018}, inverse scattering problem
\cite{Yin_2020,Meng_2020}, and structural phase transitions \cite{Landig_2015}.

To be more precise, suppose that we are interested in a real function $f(x)$ of a real
variable $x$. The function is assumed to be sign-definite. Without loss of generality,
it is sufficient to consider non-negative (positive-valued) functions. It may happen
that the most important information required for us is the tendency of the function
to infinity, where it may possess the asymptotic behavior
\vspace{12pt}
\be
\label{1}
 f(x) \simeq B x^\bt \qquad ( x \ra \infty) \;  .
\ee
In many cases, it is even not the whole function which is important, but the character of
its approach to infinity, that is, the characteristic large-variable exponent $\beta$.
However, the problem is aggravated by the complexity of the equations, defining the
function $f(x)$, to such an extent that all we are able to derive is the truncated
asymptotic expansion at small variables,
\be
\label{2}
f(x) \simeq f_k(x) \qquad ( x \ra 0) \;   ,
\ee
having the form of a finite series
\be
\label{3}
 f_k(x) = \sum_{n=0}^k a_n x^n \;  ,
\ee
where $a_0 > 0$. For simplicity, we may set $a_0 = 1$, which is equivalent to
considering the function $f(x)/a_0$.

Thus, we are put in front of the difficult task: How, knowing only the truncated
series (\ref{3}), valid for $x\ra 0$, could we extract the large-variable exponent
$\beta$ at $x \ra \infty$? As is evident, the direct application of the Pad\'{e}
summation \cite{Baker_1} of series (\ref{3}) is not applicable. This is because
the Pad\'{e} approximant $P_{M/N}(x)$ at large $x$ behaves according to the law
$x^{M-N}$, that is, it is actually not defined, as far as $M$ and $N$ can be arbitrary,
provided that $M + N = k$. The standard way of finding out the characteristic exponents
of large-variable behavior is the diff-log Pad\'{e} summation \cite{Baker_1,Guttmann_2}.
However, the question arises: How trustworthy is the result of this method? This question
is especially significant when there is no firm information of the exact value of the
sought exponent.

In such a case, the main method of proving the reliability of numerical results is
the method of validation using other solutions \cite{Awrejcewicz_2011}, when there exist
several methods of calculating the quantity of interest and all of them yield the results
compatible with each other. This technique compares the results to be validated with the
results obtained through other numerical methods. In other words, several methods to
solve the problem validate each other if the different used techniques give close results.

Thus, to make the results of calculations trustworthy, it is necessary to have several
techniques in addition to the standard diff-log Pad\'{e} summation. It is the goal of
the present paper to suggest and analyze several methods allowing us to determine the
large-variable exponent and to compare their predictions between themselves and with the
most known method of the diff-log Pad\'{e} approximation. We consider new methods
involving self-similar factor approximants. The methods to be analyzed are introduced
in Section~\ref{sec2}. These are the standard diff-log Pad\'{e} summation, the method of self-similar
factor approximants, the method combining the diff-log transformation with self-similar
factor approximants, and the approach employing Borel summation in combination with
self-similar factor approximants. In the following sections, we apply these methods
to several asymptotic series with the structure typical of many physics problems.

\section{Retrieving Large-Variable Exponents}\label{sec2}

The necessity of having several methods for finding large-variable exponents is dictated
by two reasons. First, some of the methods might be not applicable for particular cases,
which could be compensated by the use of other methods. Second, as has been explained
above, when several ways of calculating the exponents are available, it is possible to
check their consistency and, thus, to validate their use.

\subsection{Diff-Log Pad\'{e} Transformation}\label{sec2.1}

The most well known and usually employed method for defining characteristic exponents
is the diff-log Pad\'{e} transformation \cite{Baker_1,Guttmann_2}. For a given function
$f(x)$ the diff-log transformation is defined as
\be
\label{4}
 D(x) \equiv \frac{d}{dx} \; \ln f(x) \;  .
\ee
When the value of $f(x_0)$ at some point $x_0$ is known, the inverse transformation
becomes
\be
\label{5}
f(x) = f(x_0) \; \exp\left\{  \int_{x_0}^x D(t) \; dt \right\} \; .
\ee
If the function at large $x \ra \infty$ behaves as in (\ref{1}), the large-variable
exponent is given by the limit
\be
\label{6}
 \bt = \lim_{x\ra\infty} \; x D(x) \;  .
\ee

In practical applications we have, not a function, but the truncated series (\ref{3}).
Then its diff-log transform reads as
\be
\label{7}
 D_k(x) \equiv \frac{d}{dx} \; \ln f_k(x) \;  ,
\ee
where the right-hand side is expanded in powers of $x$, yielding the series
\be
\label{8}
 D_k(x) = \sum_{n=0}^k b_n x^n \qquad ( x\ra 0) \; .
\ee
The coefficients $b_n$ are uniquely defined through the coefficients $a_n$, provided
both sides of  {Equation} (\ref{7}) have the same number of terms. For expansion (\ref{8}), one
constructs the Pad\'{e} approximants
\be
\label{9}
  P_{n/n+1}[\; D_k(x) \; ] =
\frac{b_0 + \sum_{m=1}^n c_m x^m}{1+\sum_{m=1}^{n+1} d_m x^m} \; ,
\ee
with $2n + 1 = k$ and the coefficients $c_n$ and $d_n$ being defined through $b_n$.
This makes it straightforward to define the approximations for the large-variable
exponent as
\be
\label{10}
  \bt_n =  \lim_{x\ra\infty} \; x  P_{n/n+1}[\; D_k(x) \; ] \; .
\ee
As
$$
  P_{n/n+1}[\; D_k(x) \; ] \simeq \frac{c_n}{d_{n+1}} \; \left( \frac{1}{x} \right)
\qquad ( x\ra \infty) \;  ,
$$
we come to the exponents
\be
\label{11}
 \bt_n = \frac{c_n}{d_{n+1}} \;  .
\ee
This is the standard scheme for determining the characteristic exponents. By the
change of variables, the same scheme can be applied for the estimation of critical
exponents at finite values of variables.

\subsection{Self-Similar Factor Approximants}\label{sec2.2}

Recently, a new approach of extrapolating asymptotic series has been advanced
\cite{Gluzman_3,Yukalov_4,Yukalov_5,Yukalova_6} called the method of self-similar
factor approximants. This approach allows for a direct definition of characteristic
exponents by extrapolating the initial series (\ref{3}) to the form
\be
\label{12}
 f_k^*(x) = \prod_{j=1}^{N_k} (1 + A_j x)^{n_j} \;  ,
\ee
in which
\begin{eqnarray}
\label{13}
N_k = \left\{ \begin{array}{ll}
k/2 \; , ~ & ~ k = 2,4,6,\ldots \\
(k+1)/2 \; , ~ & ~ k = 1,3,5,\ldots
\end{array} \right. \; .
\end{eqnarray}
The parameters $A_j$ and $n_j$ are uniquely defined by the accuracy-through-order
procedure from equating the like-order terms in the expansions at small $x$:
\be
\label{14}
  f_k^*(x) \simeq f_k(x) \qquad ( x \ra 0 ) \; .
\ee
This procedure yields the equations
\be
\label{15}
 \sum_{j=1}^{N_k} n_j A_j^m = J_m \qquad ( m = 1,2,\ldots, k) \;  ,
\ee
with the right-hand side
$$
 J_m = \frac{(-1)^{m-1}}{(m-1)!}\; \lim_{x\ra 0} \; \frac{d^m}{d\; x^m} \;
\ln\left( 1 + \sum_{n=1}^m a_n x^n \right) \;  .
$$

Equations (\ref{15}) uniquely define all parameters $A_j$ and $n_j$ for the even
orders $k$ of expansion (\ref{3}). For the odd orders $k$, an additional normalization
condition is required which, based on scaling arguments, implies that one of the $A_j$
can be set to one \cite{Yukalova_6,Yukalov_25}. The other possibility could be by
optimizing the factor approximant (\ref{12}) with respect to one of $A_j$. Both ways
lead to close results \cite{Yukalov_7}, due to which we use the simplest
variant of setting one of $A_j$ to one. Recall that by agreement we keep in mind
real functions. Therefore form (\ref{12}) also has to be real. This requires that either
all $A_j$ are non-negative and $n_j$ real, or $A_j$ and $n_j$ can be complex, but
entering the product (\ref{12}) in complex conjugate pairs, so that their product
remains real. Occasionally arising complex-valued approximants are discarded.

At large $x$, the factor approximant (\ref{12}) results in the behavior
\be
\label{16}
 f_k^*(x) \simeq B_k x^{\bt_k} \qquad ( x \ra \infty ) \;  ,
\ee
with the amplitude
\be
\label{17}
B_k = \prod_{j=1}^{N_k} A_j^{n_j}
\ee
and the characteristic exponent
\be
\label{18}
 \bt_k = \sum_{j=1}^{N_k} n_j \;  .
\ee

\subsection{Self-Similar Diff-Log Transformation}\label{sec2.3}

As factor approximants provide an efficient tool for extrapolating asymptotic
series, it looks reasonable to try to use in the diff-log transformed expansion
(\ref{8}), instead of Pad\'{e} approximants, the self-similar factor approximants.
That is, we sum the series (\ref{8}) to the self-similar approximant
\be
\label{19}
 D_k^*(x) = b_0 \prod_{j=1}^{N_k} (1 +  L_j x)^{m_j}  \; ,
\ee
instead of the Pad\'{e} approximant (\ref{9}). As has been explained below Equation (\ref{15}),
considering real functions, the parameters $L_j$ and $n_j$ are to be such that the
approximant (\ref{19}) is real valued. 
This requires that either $L_j$ are non-negative
and $n_j$ real, or $L_j$ and $n_j$ can be complex, but entering the product (\ref{12})
in complex conjugate pairs, so that their product remains real. To make meaningful
definition (\ref{6}), the factor approximant (\ref{19}) is complemented by the condition
\be
\label{20}
 \sum_{j=1}^{N_k} m_j = -1 \;   .
\ee
Then the large-variable limit of (\ref{19}) becomes
\be
\label{21}
 D_k^*(x) \simeq D_k \; \frac{1}{x} \qquad ( x \ra\infty) \; ,
\ee
with the amplitude
\be
\label{22}
 D_k = b_0 \prod_{j=1}^{N_k}  L_j^{m_j}  \;   .
\ee
Thus the characteristic exponent
\be
\label{23}
\bt_k = \lim_{x\ra\infty} \; x D_k^*(x)
\ee
becomes
\be
\label{24}
 \bt_k = D_k = b_0 \prod_{j=1}^{N_k}  L_j^{m_j}  \;  .
\ee

\subsection{Self-Similar Borel Summation}\label{sec2.4}

The Borel transformation of the series (\ref{3}) is
\be
\label{25}
 B_k(x) = \sum_{n=0}^k \frac{a_n}{n!}\; x^n \;  .
\ee
The resulting series can be summed by means of self-similar factor approximants,
\be
\label{26}
 B_k^*(x) = \prod_{j=1}^{N_k} (1 +  M_j x)^{s_j}  \;  .
\ee
Then the sought function is approximated by the expression
\be
\label{27}
 f_k^*(x) = \int_0^\infty e^{-t} B_k^*(xt) \; dt \;  .
\ee

The Borel transform (\ref{26}) at large $x$ behaves as
\be
\label{28}
  B_k^*(x) \simeq C_k x^{\sgm_k} \qquad ( x\ra \infty) \;  ,
\ee
where
\be
\label{29}
 C_k = \prod_{j=1}^{N_k} M_j^{s_j} \; , \qquad
\sgm_k = \sum_{j=1}^{N_k} s_j \; .
\ee
Therefore, the sought function (\ref{27}), in the limit of large $x$, reduces to
\be
\label{30}
f_k^*(x) \simeq C_k x^{\sgm_k} \int_0^\infty e^{-t} t^{\sgm_k} \; dt \qquad
( x \ra \infty) \;   .
\ee
As a result, the large-variable behavior of the function acquires the form
\be
\label{31}
 f_k^*(x) \simeq B_k x^{\bt_k} \qquad ( x\ra \infty) \;  ,
\ee
with the amplitude
\be
\label{32}
B_k = C_k \Gm(\sgm_k + 1)
\ee
and the large-variable exponent
\be
\label{33}
 \bt_k = \sgm_k = \sum_{j=1}^{N_k} s_j  \; .
\ee

It is worth mentioning that the Pad\'{e} summation of the Borel transform (\ref{25})
cannot be used here. This is because, employing a Pad\'{e} approximant $P_{M/N}(x)$,
we would find for the large-variable exponent $\sigma$ the undefined value $M-N$ that,
in addition, can only be an integer.

\subsection{Simplified Self-Similar Borel Summation}\label{sec2.5}

As is shown in the previous section, the large-variable exponent of the sought
function (\ref{27}) coincides with that for the self-similar Borel transform (\ref{26}),
that is $\beta_k = \sigma_k$. Of course, the transform $B_k^*(x)$ itself is rather
different from the function $f_k^*(x)$, being connected with the latter through the
integral (\ref{27}). Instead of taking the integral, the function $f_k^*(x)$ can be
reconstructed from $B_k^*(x)$ by the method of self-similarly corrected Pad\'{e}
\mbox{approximants \cite{Gluzman_8,Gluzman_9}}. For this purpose, one can look for the
function
\be
\label{34}
  f_k^*(x) \simeq B_k^*(x) \; P_{n/n}(x) \qquad ( 2n = k ) \; ,
\ee
defining the parameters of the diagonal Pad\'{e} approximant from the
accuracy-through-order procedure by equating the like-order terms of the small-variable
expansion of (\ref{34}) and of the given expansion (\ref{3}). We have checked by several
examples and found that the so reconstructed approximation provides the accuracy
comparable to that given by the method of self-similarly corrected Pad\'{e} approximants
\cite{Gluzman_8,Gluzman_9}. The correctness of the large-variable behavior is guaranteed
by the equality (\ref{33}).

\subsection{Self-Similar Borel--Leroy Summation}\label{sec2.6}

A variant of the integral transform, slightly generalizing the Borel summation method,
is the Borel--Leroy transform that for the truncated series (\ref{3}) reads as
\be
\label{BL_1}
B_k(x,u) = \sum_{n=0}^k \frac{a_n}{\Gm(n+1+u)} \; x^n \;   ,
\ee
where $u$ plays the role of a control parameter that has to be chosen so that to
improve the convergence of the sequence of approximants, if needed. Summing up the
latter series by means of self-similar factor approximants yields
\be
\label{BL_2}
 B_k^*(x,u) =
\frac{a_0}{\Gm(1+u)} \prod_{j=1}^{N_k} (1 + A_j x)^{n_j} \; .
\ee
Accomplishing the inverse Borel--Leroy transformation gives the approximation for
the sought function
\be
\label{BL_3}
 f_k^*(x) = \int_0^\infty e^{-t} t^u  B_k^*(tx,u) \; dt \; .
\ee

At large values of the variable, the self-similar Borel--Leroy transform behaves as
\be
\label{BL_4}
 B_k^*(x,u)  \simeq C_k(u) x^{\bt_k} \qquad ( x \ra \infty) \; ,
\ee
with the amplitude
\be
\label{BL_5}
C_k(u) =  \frac{a_0}{\Gm(1+u)} \prod_{j=1}^{N_k}  A_j^{n_j}
\ee
and the exponent
\be
\label{BL_6}
\bt_k = \bt_k(u) = \sum_{j=1}^{N_k} n_j \; .
\ee
Therefore, at large values of $x$, function (\ref{BL_3}) acquires the form
\be
\label{BL_7}
f_k^*(x) \simeq B_k(u) x^{\bt_k} \qquad ( x \ra \infty) \;   ,
\ee
where the exponent $\beta_k$ is defined in (\ref{BL_6}) and the amplitude is
\be
\label{BL_8}
B_k(u) = C_k(u) \Gm( 1 + u +\bt_k) = \frac{\Gm(1+u+\bt_k)}{\Gm(1+u)} \;
a_0 \prod_{j=1}^{N_k} A_j^{n_j} \;  .
\ee

\vskip 3mm
{\bf Remark}
\vskip 2mm
 Aiming at comparing the methods of defining the large-variable
exponents, we consider several examples whose asymptotic expansions possess the
structure typical of many problems in physics and applied mathematics. Our main
aim is to study whether the methods described above can provide reasonably accurate
evaluation of the large-variable exponents for the typical cases where not so many
terms of asymptotic expansions (usually not more than about ten) are available.
This is the standard situation in the majority of physical problems of interest.
The related typical feature of the overwhelming realistic problems is that the
general expressions for the expansion coefficients are not known, hence the
convergence of the sequence of approximants cannot be checked explicitly. In such
a case, one can talk only about {\it numerical convergence} that can be observed
by comparing the numerical values of the available approximants. Under numerical
convergence, one understands the apparent approach to a limit of the given finite
sequence of numerical results. throughout this paper, we discuss only this
numerical convergence.

\section{Partition Function of Anharmonic Model}\label{sec3}

Let us start with the standard touchstone that is always considered when studying new
methods. This is the partition function of the so-called zero-dimensional anharmonic model
\be
\label{35}
Z = \frac{1}{\sqrt{\pi} }
\int_{-\infty}^\infty \exp\left( - \vp^2 - g\vp^4 \right) \; d\vp \;  ,
\ee
with the coupling parameter $g \geq 0$. Expanding the integrand in powers of $g$ leads
to the divergent series (\ref{3}) with the coefficients
\be
\label{36}
 a_n = \frac{(-1)^n}{\sqrt{\pi}\; n!} \; \Gm\left( 2n + \frac{1}{2}\right) \; .
\ee
The strong-coupling form of (\ref{35}) is
\be
\label{37}
 Z(g) \simeq 1.022765 \; g^{-0.25} \qquad ( g \ra \infty) \; .
\ee
Using the methods described above for defining the large-variable exponents $\beta_k$
for different approximants, we obtain the following results.

(i) In the case of the standard diff-log Pad\'{e} transformation of Section~\ref{sec2.1}, we have
the exponents
$$
\bt_3 = -0.1290 \; , \qquad  \bt_5 = -0.1484 \; , \qquad  \bt_7 = -0.1610 \; ,
\qquad \bt_{9} = -0.1700 \  ,
$$
numerically converging from above to the exact value $-0.25$.

(ii) Applying the self-similar factor approximants of Section~\ref{sec2.2}, we find for the even
approximants
$$
\bt_4 = -0.1290 \; , \qquad  \bt_6 = -0.1484 \; , \qquad  \bt_8 = -0.1610 \; ,
\qquad \bt_{10} = -0.1700 \; ,
$$
and for the odd orders,
$$
\bt_3 = -0.3462 \; , \qquad  \bt_5 = -0.2551 \; , \qquad  \bt_7 = -0.2227 \; ,
\qquad \bt_9 = -0.2087 \;  .
$$
One can notice monotonic numerical convergence from above for the even approximants
to the exact value $0.25$.

(iii) The self-similar diff-log transformation of Section~\ref{sec2.3} yields for the even orders
$$
 \bt_2 = -0.2281 \; , \qquad  \bt_4 = -0.1940 \; , \qquad  \bt_6 = -0.1868 \; ,
\qquad \bt_8 = -0.1853 \;  ,
$$
and the odd approximants are
$$
 \bt_3 = -0.1368 \; , \qquad  \bt_5 = -0.1586 \; , \qquad  \bt_7 = -0.1721 \; ,
\qquad \bt_9 = -0.1813 \;  .
$$
Here, the role of even and odd approximants is interchanged due to the additional
constraint (\ref{20}). Odd approximants demonstrate numerical convergence from
above.

(iv) The self-similar Borel summation of Section~\ref{sec2.4} gives for the even approximants
$$
 \bt_2 = -0.2069 \; , \quad  \bt_4 = -0.2257 \; , \quad  \bt_6 = -0.2330 \; ,
\quad \bt_8 = -0.2369 \; , \quad  \bt_{10} = -0.2394 \; ,
$$
and for the odd orders
$$
 \bt_3 = -0.2364 \; , \qquad  \bt_5 = -0.2309 \; ,  \qquad \bt_9 = -0.2419 \; .
$$
Again, the even approximants monotonically converge from above. The odd approximant for
$\beta_7$ is not defined, as it becomes complex valued.

Comparing the accuracy of the approximants, we see that the self-similar Borel summation
provides a slightly better accuracy than other methods and that the even approximants
are better than odd.

(v) It is interesting that using the self-similar Borel--Leroy summation it is possible
to find the exact value of the large-variable amplitude. To this end, let us consider
the second-order self-similar approximant for the Borel--Leroy transform (\ref{BL_2})
\be
\label{Z_1}
 B_2^*(x,u) = \frac{1}{\Gm(1+u)} \; ( 1 + A x)^{\bt_2} \;  ,
\ee
for which we have
$$
A =
\frac{35\Gm^2(2+u)-3\Gm(1+u)\Gm(3+u)}{4\Gm(2+u)\Gm(3+u)} = A(u) \; ,
$$
$$
\bt_2 =
\frac{3\Gm(1+u)\Gm(3+u)}{3\Gm(1+u)\Gm(3+u)-35\Gm^2(2+u)} = \bt_2(u) \; .
$$
The large-variable exponent $-1/4$ can be derived from the scaling relations for the
partition function (\ref{35}). Thence it should be: $\beta_2(u) = -1/4$, which results
in the control parameter $u = -0.25$. Substituting this parameter into the amplitude
(\ref{BL_8}) gives $B_2(u) = 1.02277$, which coincides with the amplitude in the
asymptotic form (\ref{37}).

\section{Quartic Anharmonic Oscillator}\label{sec4}

The other touchstone for checking new methods is the one-dimensional quartic oscillator
with the Hamiltonian in dimensionless units
\be
\label{38}
 H = - \; \frac{1}{2} \; \frac{d^2}{d\; x^2} + \frac{1}{2} \; x^2 + g x^4 \; ,
\ee
in which $x \in (-\infty, \infty)$ and $g \geq 0$. One usually calculates the
ground-state energy $E(g)$ of this oscillator.

The expansion of $E(g)$ in powers of the coupling $g$ results in a divergent series of
type (\ref{3}). The coefficients $a_n$ can be found in Refs. \cite{Bender_10,Hioe_11}.
The strong-coupling behavior is
\be
\label{39}
 E(g) \simeq 0.667986 \; g^{1/3} \qquad ( g \ra \infty) \;  .
\ee
The summary of the results obtained by different methods are as follows.

(i) Diff-log Pad\'{e} transformation (Section~\ref{sec2.1}) yields the strong-coupling exponents
$$
 \bt_3 = 0.2312 \; , \qquad  \bt_5 = 0.2570 \; , \qquad  \bt_7 = 0.2719 \; ,
\qquad \bt_9 = 0.2817 \;    .
$$
There exists monotonic numerical convergence from below to the exact value $1/3$.

(ii) Self-similar factor summation (Section~\ref{sec2.2}) gives the even approximants
$$
\bt_4 = 0.2312 \; , \qquad  \bt_6 = 0.2570 \; , \qquad  \bt_8 = 0.2719 \; ,
\qquad \bt_{10} = 0.2817 \;
$$
and the odd approximants
$$
\bt_3 = 0.5903 \; , \qquad  \bt_5 = 0.4092 \; , \qquad  \bt_7 = 0.3510 \; ,
\qquad \bt_9 = 0.3276 \; .
$$
Even approximants monotonically converge from below.

(iii) Self-similar diff-log transformation (Section~\ref{sec2.3}) results in the even approximants
$$
\bt_2 = 0.3803 \; , \qquad  \bt_4 = 0.3170 \; , \qquad  \bt_6 = 0.3033 \; ,
\qquad \bt_8 = 0.2996 \; ,
$$
and in the odd approximants
$$
\bt_3 = 0.2408 \; , \qquad  \bt_5 = 0.2674 \; , \qquad  \bt_7 = 0.2818 \; ,
\qquad \bt_9 = 0.2909 \;  .
$$
Again, we have to remember that here, due to the constraint (\ref{20}), the role of
the even and odd approximants is interchanged. Here, the odd approximants monotonically
converge from below.

(iv) Self-similar Borel summation (Section~\ref{sec2.4}) leads to the even approximants
$$
\bt_2 = 0.3 \; , \qquad  \bt_4 = 0.2891 \; , \qquad  \bt_6 = 0.3119 \; ,
\qquad \bt_{10} = 0.3219 \; ,
$$
and to the odd approximants
$$
\bt_3 = 0.2368 \; , \qquad  \bt_5 = 0.3305 \; , \qquad  \bt_7 = 0.3147 \; ,
\qquad \bt_9 = 0.3203 \;    .
$$
Even approximants converge from below to the exact value $1/3$.

\section{Expansion Factor of Polymer Chain}\label{sec5}

The theory of the excluded volume effect in a polymer chain has been one of the central
problems in the field of polymer solution theory. The net effect of the excluded volume
interaction between segments of the polymer chain is usually repulsive and leads to an
expansion of the chain size. There have been many attempts to understand, quantitatively,
this effect over the past several decades \cite{Muthukumar_12,Muthukumar_13}. When the
excluded volume interaction is very weak, a perturbation theory for the ratio of the mean
square end-to-end distance of the chain to its unperturbed value can be developed and
can be reduced to a series in a single dimensionless interaction parameter $g$. This ratio,
called expansion factor $\alpha(g)$, derived by means of perturbation theory
\cite{Muthukumar_12,Muthukumar_13} with respect to the coupling parameter $g$, results
in a series (\ref{3}) with the coefficients
$$
a_0 = 1 \; , \qquad a_1 = \frac{4}{3} \; , \qquad a_2 = -2.075385396 \; \qquad
a_3 = 6.296879676 \; ,
$$
$$
a_4 = -25.05725072 \; \qquad a_5 = 116.134785 \; ,  \qquad
a_6 = -594.71663 \; .
$$
The strong-coupling behavior has been found numerically \cite{Li_14} in the form
\be
\label{41}
\al(g) \simeq 1.5309 \; x^{0.3544} \;  .
\ee
The following results are obtained.

(i) Diff-log Pad\'{e} transformation (Section~\ref{sec2.1}) leads to
$$
\bt_3 = 0.3400 \; , \qquad  \bt_5 = 0.3477 \; ,
$$
These values approach the exponent $0.3544$ from below.

(ii) Self-similar factor approximants (Section~\ref{sec2.2}) give in even orders
$$
\bt_4 = 0.3400 \; , \qquad \bt_6 = 0.3477 \;    .
$$
and in odd orders
$$
\bt_3 = 0.4399 \; , \qquad  \bt_5 = 0.3641 \;    .
$$
Even approximants are closer to the test value $0.3544$.

(iii) Self-similar diff-log transformation (Section~\ref{sec2.3}) provides in even orders
$$
\bt_2 = 0.3795 \; , \qquad \bt_4 = 0.3542 \; ,
$$
and in odd orders
$$
\bt_3 = 0.3430 \; , \qquad  \bt_5 = 0.3488 \; .
$$
All these values are close to each other.

(iv) In the case of the self-similar Borel summation (Section~\ref{sec2.4}), we find the even
approximants
$$
\bt_2 = 0.4614 \; , \qquad \bt_4 = 0.3184 \; \qquad \bt_6 = 0.3726 \; ,
$$
and the odd approximants
$$
\bt_3 = 0.2417 \; , \qquad  \bt_5 = 0.4489 \;    .
$$

\section{Massive Schwinger Model}\label{sec6}

The massive Schwinger model in Hamiltonian lattice theory \cite{Schwinger_15,Hamer_16}
describes quantum electrodynamics in two space-time dimensions. Its features include
such properties of quantum chromodynamics as confinement, chiral symmetry breaking, and
a topological vacuum. Due to this, the model has attracted much attention. It is
perhaps the simplest non-trivial gauge theory, and this makes it a standard test-bed
for the trial of new techniques for the studies. The main characteristic of interest
in the Schwinger model is the spectrum of bound states, more specifically the lowest
two bound states and the energy gap between them that can be calculated perturbatively.

Let us consider the energy gap between the lowest and first excited states of the vector
boson as a function $\Delta(z)$ of the variable $z = (1/ga)^4$, where $g$ is a coupling
parameter and $a$, lattice spacing. This energy gap at small $z$ can be represented
as a series
\be
\label{42}
\Dlt(z) \simeq \sum_n a_n z^n \qquad ( z \ra 0 ) \;  ,
\ee
with the coefficients
$$
a_0 = 1 \; , \qquad a_1 = 2 \; , \qquad a_2 = -10 \; \qquad
a_3 = 78.66667 \; , \qquad a_4 = -736.2222 \; ,
$$
$$
a_5 = 7572.929 \; , \qquad a_6 = -{82{,}736}.69 \; , \qquad a_7 = 942{,}803.4 \; ,
$$
\be
\label{43}
a_8 = -1.108358 \times 10^7  \; ,  \qquad a_9 = 1.334636 \times 10^8 \; ,
\qquad a_{10} =-1.637996 \times 10^9 \; .
\ee
In the continuous limit, where the lattice spacing tends to zero, the variable $z$ tends
to infinity. Then the gap acquires the limiting form
\be
\label{44}
\Dlt(z) \simeq  0.5642 \; z^{1/4} \qquad ( z \ra \infty)  \;  .
\ee

(i) Using the diff-log Pad\'{e} transformation (Section~\ref{sec2.1}), we find the large-variable
exponents
$$
\bt_3 = 0.1845 \; , \qquad  \bt_5 = 0.1933 \; , \qquad  \bt_7 = 0.1983 \; ,
\qquad  \bt_9 = 0.2023 \; ,
$$
which are below the value $0.25$.

(ii) Employing self-similar factor approximants (Section~\ref{sec2.2}), we have the even
approximants
$$
\bt_4 = 0.1845 \; , \qquad  \bt_6 = 0.1933 \; , \qquad  \bt_8 = 0.1983 \; ,
\qquad  \bt_{10} = 0.20234 \; ,
$$
and the odd approximants
$$
\bt_3 = 0.2714 \; , \qquad  \bt_5 = 0.2140 \; , \qquad \bt_7 = 0.2036 \; ,
\qquad  \bt_9 = 0.20233 \; .
$$

(iii) Self-similar diff-log transformation (Section~\ref{sec2.3}) produces the even approximants
$$
\bt_2 = 0.2014 \; , \qquad  \bt_4 = 0.1970 \; ,
$$
and the odd approximants
$$
\bt_3 = 0.1882 \; , \qquad  \bt_5 = 0.1991 \; .
$$
The higher-order approximants are discarded, being complex-valued.

(iv) Making use of the self-similar Borel summation (Section~\ref{sec2.4}) gives oscillating even
approximants
$$
\bt_2 = 0.2857 \; , \qquad \bt_4 = 0.1230 \; , \qquad \bt_6 = 0.2643 \; ,
\qquad \bt_8 = 0.1547 \; , \qquad \bt_{10} = 0.2496 \; ,
$$
while the odd approximants oscillate so widely that they lose their meaning.

One often assumes that the Borel method improves the results of series summation.
However, it is necessary to be cautious. Thus, the considered above Schwinger model shows
that Borel summation can produce nonmonotonic sequences of approximants, as compared
to the direct self-similar summation of the given series.

\section{Equation of State for Hard-Disc Fluid}\label{sec7}

The fluid of hard discs of diameter $a_s$, which approximately equals the scattering
length for these objects, is an important model often used as a realistic approximation
for systems with more complicated interaction potentials. The equation of state connects
pressure $P$, temperature $T$, and density $\rho$. One often considers the ratio
\be
\label{F_1}
 Z = \frac{P}{\rho T} \; ,
\ee
where the Planck constant is set to one, which is called compressibility factor. This
factor is studied as a function of the packing fraction, or filling,
\be
\label{F_2}
f \equiv \frac{\pi}{4} \;\rho\; a_s^2  \;  .
\ee
It is known \cite{Mulero_21,Santos_22} that the compressibility factor exhibits
critical behavior at the filling $f_c = 1$, where
\be
\label{F_3}
 Z ~ \propto ~ ( f_c - f)^{-\al} \qquad ( f \ra f_c -0 ) \; .
\ee
The exponent $\alpha$ has not been calculated exactly, but it is conjectured
\cite{Mulero_21,Santos_22} to be around $\alpha = 2$.

The compressibility factor for low-density has been found \cite{Clisby_23,Maestre_24}
by perturbation theory as an expansion in powers of the filling $f$. Nine terms of
this expansion are available:
$$
Z \simeq 1 + 2f + 3.12802 f^3 + 4.25785 f^3 + 5.3369 f^4 + 6.36296 f^5 +
$$
\be
\label{F_4}
 + 7.35186 f^6 +
8.3191 f^7 + 9.27215 f^8 + 10.2163 f^9 \;  ,
\ee
where $f \ra 0$. In order to reduce the consideration to the same type of problems
as treated above, we make the substitution
\be
\label{F_5}
f = \frac{x}{1+x} \; f_c = \frac{x}{1+x} \; ,
\qquad
x = \frac{f}{f_c-f} = \frac{f}{1-f} \;  .
\ee
Then
\be
\label{F_6}
 x \ra 0 \qquad ( f\ra 0 ) \; ,
\qquad
 x \ra \infty \qquad ( f\ra f_c-0 )  \; ,
\ee
and the compressibility factor at the critical point behaves as
\be
\label{F_7}
Z ~ \propto ~ x^\al \qquad ( x \ra \infty) \; .
\ee
With substitution (\ref{F_5}), the compressibility factor $Z$, as a function of $x$
becomes
$$
Z \simeq 1 + 2x + 1.12802 x^2 + 0.00181 x^3 - 0.05259 x^4 +
$$
\be
\label{F_8}
  + 0.05038 x^5 - 0.03234 x^6
+ 0.01397 x^7 - 0.0033 x^8 + 0.00618 x^9 \;   ,
\ee
where $x \ra 0$. Thus the problem reduces to the prediction of the large-variable
exponent $\alpha$, being based on the small-variable expansion (\ref{F_8}).

(i) Using the standard diff-log Pad\'{e} transformation, we have
$$
\al_3 = 1.6186 \; , \qquad \al_5 = 1.8498 \; , \qquad \al_7 = 1.8663 \; ,
\qquad \al_9 = 3.2337 \;  ,
$$
with the approximation values increasing above $3$.

(ii) Self-similar factor approximants give in even orders
$$
 \al_4 = 1.6186 \; , \qquad \al_6 = 1.8498 \; , \qquad \al_8 = 1.8663 \; ,
$$
and in odd orders,
$$
\al_3 = 2.1294 \;  , \qquad \al_5 = 1.8432 \; , \qquad \al_7 = 1.8400 \; .
$$
The even approximants increase, while the odd approximants decrease. Again we see
that the diff-log Pad\'{e} approximants of order $k$ coincide with the self-similar
approximants of order $k+1$.

(iii) Self-similar diff-log transformation results in the even approximants
$$
\al_2 = 2.1432 \;  , \qquad \al_4 = 1.8727 \; , \qquad \al_6 = 1.8546 \; ,
$$
and odd approximants
$$
\al_3 = 1.6995 \;  , \qquad \al_5 = 1.8478 \; , \qquad \al_7 = 1.8628 \; .
$$
These values are close to the exponents obtained by direct self-similar summation
of series (\ref{F_8}), without the diff-log transformation.

(iv) Self-similar Borel summation leads to the even approximants
$$
\al_2 = 1.3928 \;  , \qquad \al_4 = 1.9772 \; , \qquad \al_6 = 1.6042 \; ,
\qquad \al_8 = 1.8890 \; ,
$$
and to the odd approximants
$$
\al_3 = 1.5642 \;  , \qquad \al_5 = 1.9476 \; , \qquad \al_7 = 1.6660 \;
\qquad \al_9 = 1.9058\;  .
$$
As is seen, these values are close to $2$.

(v) Self-similar Borel--Leroy summation requires to be explained in a bit more
details. We follow Section~\ref{sec2.6}, except that now, instead of the exponent $\beta_k$
we write $\alpha_k$. All other steps are the same. For the series (\ref{F_8}),
we define the Borel--Leroy transform (\ref{BL_1}), construct the self-similar
approximant (\ref{BL_2}), make the inverse transformation (\ref{BL_3}), consider
the large-variable limit (\ref{BL_4}), and find the exponent (\ref{BL_6}) that
now reads as $\alpha_k(u)$. The control parameter $u$ can be defined from the
optimization conditions \cite{Yukalov_7,Yukalov_26,Yukalov_17,Gluzman_27}. The
minimal derivative condition is spoiled by multiple solutions, while the minimal
difference condition in the form $\alpha_2(u) = \alpha_3(u)$, yields the unique
solution $u = 0.35165$. Using this control parameter, we obtain in even orders
$$
 \al_2 = 1.4797 \;  , \qquad \al_4 = 1.9700 \; , \qquad \al_6 = 1.6614 \; ,
\qquad \al_8 = 1.9164 \; ,
$$
and in odd orders
$$
\al_3 = 1.4797 \;  , \qquad \al_5 = 1.9431 \; , \qquad \al_7 = 1.6996 \;
\qquad \al_9 = 1.9290 \;   .
$$
Considering the largest-order self-similar approximants, depending on the used method,
we have in the even orders $1.866$, $1.855$, $1.889$, and $1.916$, which gives the
average $1.882$. In the odd orders, we have $1.840$, $1.863$, $1.906$, and $1.929$,
which results in the average $1.885$. Thus, we can make the prediction for the exponent
as $1.884 \pm 0.02$.

\section{Discussion}\label{sec8}

The problem of defining large-variable characteristic exponents is considered. We
analyze and compare different methods: the standard approach using diff-log Pad\'{e}
transformation, and several novel methods involving the use of self-similar
factor approximants. It is worth noting that the form of the self-similar approximants
is not postulated ad hoc, but follows from self-similar approximation theory, where
these approximants represent fixed points of renormalization group equations
\cite{Yukalov_17}.

The suggested methods are developed for the application to difficult problems
characterized by three features. First, the number of terms in an asymptotic expansion
is not large, often containing just a few terms. Second, the general explicit expression
for expansion coefficients is not available, hence it is not known what the properties
of the sought function are. Third, the expansion variable is not small, but rather large and
even tending to infinity. In such a situation, the sole known way of estimating the
efficiency of a summation method is based on (i) the observation of apparent numerical
convergence (that should not be confused with convergence in strict sense) and (ii) the
verification of the compatibility of the results obtained by several available methods.
The developed methods are applicable to any series and the results are trustful provided
they satisfy the above requirements. The methods are straightforward and do not involve
any fitting parameters.

One should not confuse the calculation of the large-variable exponents, where the
variable of interest tends to infinity, with the calculation of critical exponents
for which the extrapolation to only finite values of variables is required, as for
instance in the problem of summation of epsilon expansions, where at the end one sets
$\varepsilon = 1$. Thus, rather precise values for critical exponents have been found by
means of self-similar approximants \cite{Yukalov_17,Yukalov_18}, agreeing well with
other methods of summation and Mote Carlo simulations summarized in Refs.
\cite{Pelissetto_19,Dupuis_20}. The extrapolation of asymptotic series to the values
of the variable tending to infinity is a more complicated task even for simple problems.

In order to decide on the accuracy of the used methods, it is possible to compare
the upper-order results for each considered case, obtained by different methods. For
compactness, we shall denote the methods by the corresponding abbreviations: Diff-Log
Pad\'{e} transformation (DLP); Self-Similar Factor approximants (SSF); Self-Similar
Diff-Log transformation (SSDL); and Self-Similar Borel summation (SSB).

For the anharmonic model of Section~\ref{sec3}, we have:
$$
\beta_9 = -0.1700 {\text{~(DLP)}} \; , \qquad \beta_{10} = -0.1700 \text{~(SSF)} \; , 
\qquad \beta_9 = -0.1813 \text{~(SSDL)} \; , 
$$
$$
\qquad \beta_{10} = -0.2394 \text{~(SSB)} \; .
$$
The result of the SSB is the closest to the exact value $\beta = -0.25$.

In the case of the anharmonic oscillator of Section~\ref{sec4} the results are:
$$
\beta_9 = 0.2817 \text{~(DLP)} \; , \qquad \beta_{10} = 0.2817 \text{~(SSF)} \; , 
\qquad
\beta_9 = 0.2909 \text{~(SSDL)} \; , 
$$
$$
 \beta_{10} = 0.3219 \text{~(SSB)} \; .
$$
The value given by SSB is the closest to the exact $\beta = 1/3$.

For the polymer chain of Section~\ref{sec5}, we find:
$$
\beta_5 = 0.3477 \text{~(DLP)} \; , \qquad \beta_{6} = 0.3477 \text{~(SSF)} \; , 
\qquad
\beta_5 = 0.3488 \text{~(SSDL)} \; , 
$$
$$
\beta_{6} = 0.3726 \text{~(SSB)} \; .
$$
The closest to the exact $\beta = 0.3544$ is the result of SSDL.

For the massive Schwinger model of Section~\ref{sec6}, we find:
$$
\beta_9 = 0.2023 \text{~(DLP)} \; , \qquad \beta_{10} = 0.2023 \text{~(SSF)} \; , 
\qquad
\beta_9 = 0.1991 \text{~(SSDL)} \; , 
$$
$$
\beta_{10} = 0.2496 \text{~(SSB)} \; .
$$
The result of SSB is the closest to the exact $\beta = 0.25$.

In the case of the hard-disc fluid of Section~\ref{sec7}, we obtain:
$$
\alpha_9 = 3.2337 \text{~(DLP)} \; , \qquad \alpha_{8} = 1.8663 \text{~(SSF)} \; , 
\qquad
\alpha_7 = 1.8628 \text{~(SSDL)} \; , 
$$
$$
\alpha_{9} = 1.9058 \text{~(SSB)} \; .
$$
The closest to the conjectured $\alpha = 2$ is the result of SSB. The use of the
self-similar Borel--Leroy transformation slightly improves the result giving
$\alpha_9 = 1.9290$.

The main conclusions are as follows:

(i) It turns out that the direct summation of asymptotic series by means of self-similar
factor approximants gives the results coinciding with the standard method of diff-log
Pad\'{e} transformation. Moreover, the large-variable exponents $\beta_k$ of the latter
method coincide with the exponents $\beta_{k+1}$ of the first method. This can be
explained by the fact that Pad\'{e} approximants are just a particular case of factor
approximants.

(ii) In the methods, employing self-similar factor approximants, even approximants
demonstrate better numerical convergence than odd approximants. The reason for that is
the use for the odd approximants of an additional normalization constraint. The general
feature of the self-similar factor approximants is their self-organized structure
prescribed by renormalization-group procedure \cite{Yukalov_17}. Therefore, usually, the
lesser imposed constraints, the better the numerical convergence of the approximants.

(iii) The cases where the self-similar Borel summation is well defined lead to
more accurate results. However, sometimes it may produce strongly oscillating sequences
of the approximants, and even may stop existing in the real-valued range.

(iv) All results are compatible with each other, which validates their use. This is a
principal point, as in order to obtain reliable estimates of calcualtions, it is
necessary to have in hands several methods demonstrating the compatibility of results
between the different techniques. This is why the novel methods, considered in the
present paper, are of high importance, as they provide the tool for checking the
compatibility between different approaches, hence they demonstrate the reliability of
the obtained results.

As is possible to conclude from the comparison of different approaches, the method
of self-similar factor approximants is comparable in accuracy with the method of
diff-log transformation, while the self-similar Borel summation can provide more
accurate results.


\vskip 3cm

\begin{thebibliography}{999}
\bibitem{Sornette_2003}
Sornette, D.
{\it Critical Phenomena in Natural Sciences};
Springer: Berlin, Germany, 2003.

\bibitem{Pareto_1927}
Pareto, V.
{\it Manuel d'Economie Politique};
Girard: Paris, France, 1927.

\bibitem{Saichev_2010}
Saichev, A.; Malevergne, Y.; Sornette, D.
{\it Theory of Zipf's Law and Beyond};
Springer: Berlin, Germany, 2010.

\bibitem{Migdal_2018}
Migdal, A.B.
{\it Qualitative Methods in Quantum Theory};
CRC Press: Boca Raton, FL, USA, 2018.

\bibitem{Yin_2020}
Yin, W.; Yang, W.; Liu, H.
A neural network scheme for recovering scattering obstacles with limited phaseless
far-field data.
{\it J. Comput. Phys.} {\bf 2020}, {\it 417}, 1--26.

\bibitem{Meng_2020}
Meng P., Su, L.; Yin, W.
Solving a kind of inverse scattering problem of acoustic waves based on linear
sampling method and neural network.
{\it Alex. Eng. J.} {\bf 2020}, {\it 59}, 1451--1462.

\bibitem{Landig_2015}
Landig, R.; Brennecke, F.; Mottl, R.; Donner, T.; Essliger, T.
Measuring the dynamic structure factor of a quantum gas undergoing a structural phase
transition.
{\it Nat. Commun.} {\bf 2015}, {\it 6}, 7046.

\bibitem{Baker_1}
Baker, G.A.
Application of the Pad\'{e} approximant method to the investigation of some
magnetic properties of the Ising model.
{\it Phys. Rev.} {\bf 1961}, {\it 124}, 768--774.

\bibitem{Guttmann_2}
Guttmann, A.J.
Analysis of series expansions for nonalgebraic singularities.
{\it J. Phys. A} {\bf 2015}, {\it 48}, 045209.

\bibitem{Awrejcewicz_2011}
Awrejcewicz, J.
{\it Numerical Analysis: Theory and Applications};
Intech: Janeza Trdine, Rijeka, 2011.


\bibitem{Gluzman_3}
Gluzman, S.; Yukalov, V.I.; Sornette, D.
Self-similar factor approximants.
{\it Phys. Rev. E} {\bf 2003}, {\it 67}, 026109.

\bibitem{Yukalov_4}
Yukalov, V.I.; Gluzman, S.; Sornette, D.
Summation of power series by self-similar factor approximants.
{\it Physica A} {\bf 2003}, {\it 328}, 409--438.

\bibitem{Yukalov_5}
Yukalov, V.I.; Gluzman, S.
Extrapolation of power series by self-similar factor and root approximants.
{\it Int. J. Mod. Phys. B} {\bf 2004}, {\it 18}, 3027--3046.

\bibitem{Yukalova_6}
Yukalova, E.P.; Yukalov, V.I.; Gluzman, S.
Self-similar factor approximants for evolution equations and boundary-value problems.
{\it Ann. Phys. (N. Y.)} {\bf 2008}, {\it 323}, 3074--3090.

\bibitem{Yukalov_25}
Yukalov, V.I.; Yukalova, E.P.
From asymptotic series to self-similar approximants.
{\it Physics} {\bf 2021}, {\it 3}, 829--878.

\bibitem{Yukalov_7}
Yukalov, V.I.; Gluzman, S.
Optimization of self-similar factor approximants.
{\it Mol. Phys.} {\bf 2009}, {\it 107}, 2237--2244.

\bibitem{Gluzman_8}
Gluzman, S.; Yukalov, V.I.
Self-similarly corrected Pad\'{e} approximants for the indeterminate problem.
{\it Eur. Phys. J. Plus} {\bf 2016}, {\it 131}, 340.

\bibitem{Gluzman_9}
Gluzman, S.; Yukalov, V.I.
Self-similarly corrected Pad\'{e} approximants for nonlinear equations.
{\it Int. J. Mod. Phys. B} {\bf 2019}, {\it 33}, 1950353.

\bibitem{Bender_10}
Bender, C.M.; Wu, T.T.
Anharmonic oscillator.
{\it Phys. Rev.} {\bf 1969}, {\it 184}, 1231--1260.

\bibitem{Hioe_11}
Hioe, F.T.; MacMillen, D.; Montroll, E.W.
Quantum theory of anharmonic oscillators: Energy levels of a single and a pair of
coupled oscillators with quartic coupling.
{\it Phys. Rep.} {\bf 1978}, {\it 43}, 305--335.

\bibitem{Muthukumar_12}
Muthukumar, M.; Nickel, B.G.
Perturbation theory for a polymer chain with excluded volume interaction.
{\it J. Chem. Phys.} {\bf 1984}, {\it 80}, 5839--5850.

\bibitem{Muthukumar_13}
Muthukumar, M.; Nickel, B.G.
Expansion of a polymer chain with excluded volume interaction.
{\it J. Chem. Phys.} {\bf 1987}, {\it 86}, 460--476.

\bibitem{Li_14}
Li, B.; Madras, N.; Sokal, A.D.
Critical exponents, hyperscaling, and universal amplitude ratios for two- and
three-dimensional self-avoiding walks.
{\it J. Stat. Phys.} {\bf 1995}, {\it 80}, 661--754.

\bibitem{Schwinger_15}
Schwinger, J.
Gauge invariance and mass.
{\it Phys. Rev.} {\bf 1962}, {\it 128}, 2425--2428.

\bibitem{Hamer_16}
Hamer, C.J.; Weihong, Z.; Oitmaa, J.
Series expansions for the massive Schwinger model in Hamiltonian lattice theory.
{\it Phys. Rev. D} {\bf 1997}, {\it 56}, 55--67.

\bibitem{Mulero_21}
Mulero, A.; Cachadina, I.; Solana, J.R.
The equation of state of the hard-disc fluid revisited.
{\it Mol. Phys.} {\bf 2009}, {\it 107}, 1457--1465.

\bibitem{Santos_22}
Santos, A.; Lopez de Haro, M.; Bravo Yuste, S.
An accurate and simple equation of state for hard disks.
{\it J. Chem. Phys.} {\bf 1995}, {\it 103}, 4622--4625.

\bibitem{Clisby_23}
Clisby, N.; McCoy, B.M.
Ninth and tenth order virial coefficients for hard spheres in $D$ dimensions.
{\it J. Stat. Phys.} {\bf 2006}, {\it 122}, 15--57.

\bibitem{Maestre_24}
Maestre, M.A.G.; Santos, A.; Robles, M.; Lopez de Haro, M.
On the relation between coefficients and the close-packing of hard
disks and hard spheres.
{\it J. Chem. Phys.} {\bf 2011}, {\it 134}, 084502.

\bibitem{Yukalov_26}
Yukalov, V.I.; Gluzman, S.
Critical indices as limits of control functions.
{\it Phys. Rev. Lett.} {\bf 1997}, {\it 79}, 333--336.

\bibitem{Yukalov_17}
Yukalov, V.I.
Interplay between approximation theory and renormalization group.
{\it Phys. Part. Nucl.} {\bf 2019}, {\it 50}, 141--209.

\bibitem{Gluzman_27}
Gluzman, S.
Optimized factor approximants and critical index.
{\it Symmetry} {\bf 2021}, {\it 13}, 903.

\bibitem{Yukalov_18}
Yukalov, V.I.; Yukalova, E.P.
Calculation of critical exponents by self-similar factor approximants.
{\it Eur. Phys. J. B} {\bf 2007}, {\it 55}, 93--99.

\bibitem{Pelissetto_19}
Pelissetto, A.; Vicari, E.
Critical phenomena and renormalization-group theory.
{\it Phys. Rep.} {\bf 2002}, {\it 368}, 549--727.

\bibitem{Dupuis_20}
Dupuis, N.; Canet, L.; Eichhorn, A.; Metzner, W.; Pawlowski, J.M.; Tissier, M.; Wschebor, N.
The nonperturbative functional renormalization group and its applications.
{\it Phys. Rep.} {\bf 2021}, {\it 910}, 1--114.

\end{thebibliography}
\end{document}